# Two-level system noise reduction for Microwave Kinetic Inductance Detectors

Omid Noroozian*, Jiansong Gao†, Jonas Zmuidzinas*, Henry G. LeDuc¶, Benjamin A. Mazin§

* *Cahill Center for Astronomy and Astrophysics, California Institute of Technology, Pasadena, CA 91125, USA*
† *National Institute of Standards and Technology, Boulder, CO 80305, USA*
¶ *Jet Propulsion Laboratory, 4800 Oak Grove Drive, Pasadena, CA 91109, USA*
§ *University of California, Santa Barbara, CA 93106, USA*

**Abstract.** Noise performance is one of the most crucial aspects of any detector. Superconducting Microwave Kinetic Inductance Detectors (MKIDs) have an "excess" frequency noise that shows up as a small time dependent jitter of the resonance frequency characterized by the frequency noise power spectrum measured in units of $Hz^2/Hz$. Recent studies have shown that this noise almost certainly originates from a surface layer of two-level system (TLS) defects on the metallization or substrate. Fluctuation of these TLSs introduces noise in the resonator due to coupling of the TLS electric dipole moments to the resonator's electric field. Motivated by a semi-empirical quantitative theory of this noise mechanism, we have designed and tested new resonator geometries in which the high-field "capacitive" portion of the CPW resonator is replaced by an interdigitated capacitor (IDC) structure with $10 - 20$ $\mu$m electrode spacing, as compared to the 2 $\mu$m spacing used for our more conventional CPW resonators. Measurements show that this new IDC design has dramatically lower TLS noise, currently by about a factor of ~29 in terms of the frequency noise power spectrum, corresponding to an improvement of about a factor of $\sqrt{29}$ in NEP. These new devices are replacing the CPW resonators in our next design iteration in progress for MKIDCam. Opportunities and prospects for future reduction of the TLS noise will be discussed.

## INTRODUCTION

Detector sensitivity is the key for major advances in astronomy, especially in the submillimeter regime where only recently have sensitive and reliable data been collected. Bolometers have been the dominating technology for continuum detection in the submillimeter both in space and on ground. However, due to their limitation on array size arising from complexity in multiplexing, they are almost impractical for arrays >10000 pixels. A promising alternative are Microwave Kinetic Inductance Detectors (MKID) [1,2,3] which are inherently multiplexible to very large arrays. MKIDs can be made from simple coplanar waveguide (CPW) resonators fabricated using a single patterned superconducting thin film on a high-quality crystalline substrate. They are basically pair-breaking detectors, in which the absorbed photon energy breaks Cooper pairs inside the superconductor and creates extra quasi-particles ($N_{qp}$). These recombine with a time constant $\tau_{qp} \approx 10^{-3}$-$10^{-6}$ seconds. The film is cooled to $T \ll T_c$ and is partly made of Aluminum with a superconducting energy gap of $\Delta_0 \sim 0.18$ meV, making the detector sensitive to photons above the gap frequency of ~ 90 GHz. The surface impedance ($Z_s = R_s + j\omega L_s$) of the film depends on the quasi-particle density, and very sensitive measurements of $\delta Z_s$ can be made using a resonant coplanar-waveguide (CPW) circuit weakly coupled to a CPW feedline [1]. Changes in $L_s$ and $R_s$ affect the resonance frequency ($f_r$) and quality factor ($Q_r$) of the resonator which can be detected using a homodyne technique to read out the complex transmission ($S_{21}$) of a microwave probe signal through the feedline. These CPW resonators have shown excess frequency noise [1,2,3,4] that often limits the detectors sensitivity. The origin of this noise has been extensively studied and characterized, and has been associated with two-level systems (TLS) [4,5,6,7]. Here we present a new resonator geometry

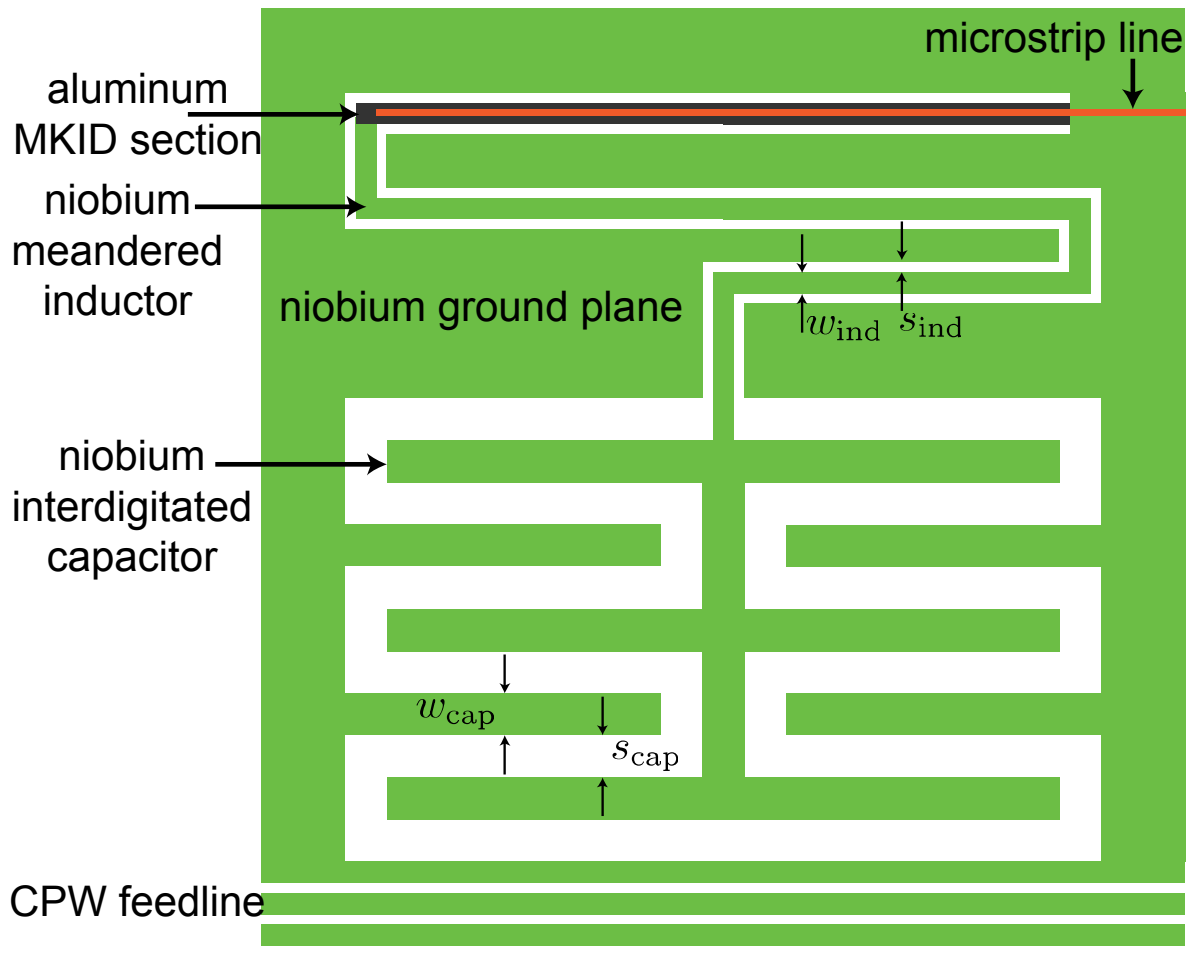

**FIGURE 1.** (Color online) A schematic illustration of an IDC resonator. The resonator is capacitively coupled to a ~50 Ω feedline. The capacitor has 3 fingers on each side. The inductor is short circuited at the aluminum end where the mm/submm radiation is absorbed.

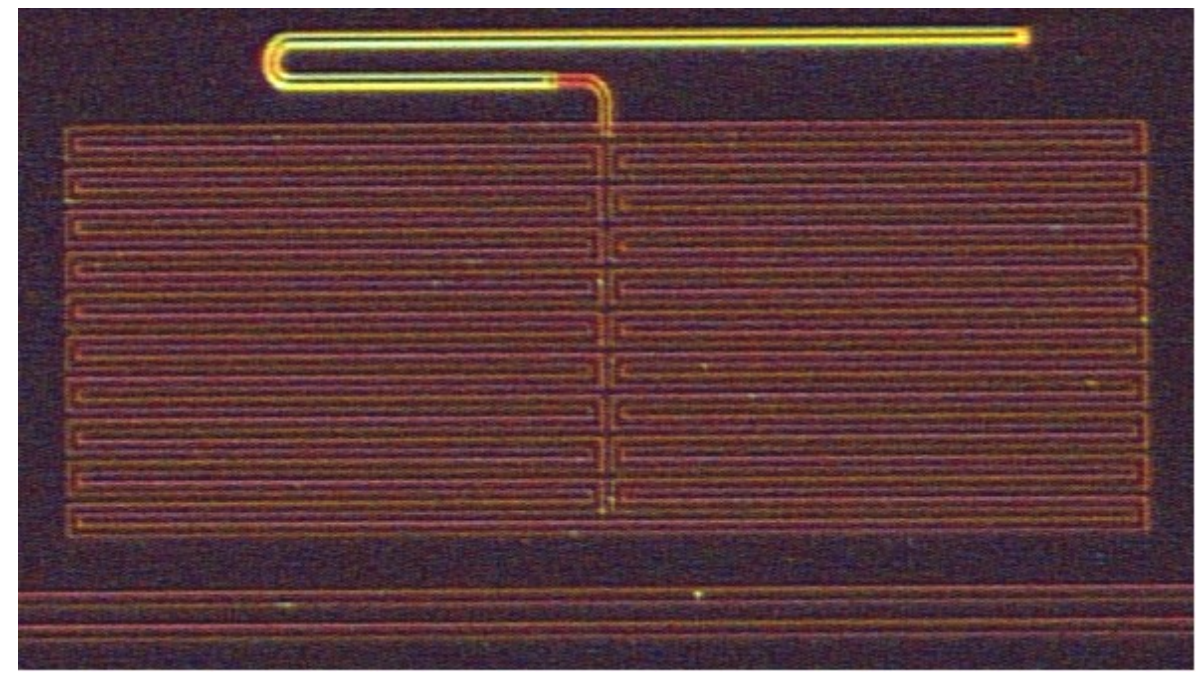

**FIGURE 2.** (Color online) Fabricated IDC resonator with 19 fingers on each side of the capacitor. The capacitor area on chip is 0.7 mm x 1 mm. Other parameters are as follows: $w_{cap} = s_{cap} = 10\mu$m, $w_{ind} = 6\mu$m, $s_{ind} = 2\mu$m, $Q_r$=145,000, $Q_c$=170,000, $f_r$=5.56 GHz, $\alpha$=4.7%, $\Delta_0$=0.186 meV.

that significantly reduces the excess noise without sacrificing the responsivity of the detector.

## MKID FREQUENCY NOISE

MKIDs show "excess" frequency noise [1,4]. The detector frequency noise is a small time-dependent jitter of the resonance frequency $f_r$ characterized by the frequency noise power spectrum $S_{\delta f_r}$ measured in units of Hz$^2$/Hz. It has been demonstrated [3,5,6] that this noise originates primarily from a surface layer on the device. This layer contains two-level defects such as are commonly found in amorphous dielectric materials, and the fluctuation of these TLSs introduces noise in the resonator due to coupling of the TLS electric dipole moments to the resonator's electric field. It has been observed [4,6] that in CPW resonators this noise scales as $P_{int}^{-1/2}$ where $P_{int}$ is the power inside the resonator, and also scales as $w^{-1.58}$ where $w$ is the center strip width of the CPW. A semi-empirical quantitative theory of this noise mechanism has been developed [6] that accurately reproduces both the observed power and geometrical scalings by assuming that the TLS noise contributions scale as the cube of the electric field ($|E|^3$). This implies that those TLS located near the open-circuit (capacitive) end of the resonator are the most harmful. This insight has been used to develop a modified resonator geometry in which the capacitive section has been replaced by an interdigitated capacitor (IDC) with wide fingers spacing to benefit from noise reduction, while the inductive low-$|E|$ end is kept narrow to maintain high kinetic inductance fraction (α) [8] and responsivity. The design and measured noise performance of this new device are described below.

## INTERDIGITATED CAPACITOR (IDC) RESONATORS

A schematic design of a resonator with an IDC is shown in Fig 1. One of the several fabricated devices is shown in Fig 2. The resonator is patterned from a 200 nm thick Nb ($T_c = 9.2$ K) film on a 450 $\mu$m crystalline Si substrate (with a thin native oxide layer expected to be present due to air exposure) using a photoresist mask and an $SF_6$ plasma etch. The IDC is made of 19 fingers on both sides where each finger is either 10 $\mu$m or 20 $\mu$m wide with either 10 $\mu$m or 20 $\mu$m spacing between them respectively. The measurement results in this paper are for the 10 $\mu$m width and gap device. The distance between the IDC and the feedline as well as the finger length controls the coupling quality factor ($Q_c$) of the resonator. The narrow meander section acts as a CPW inductor that is short circuited on one end where the mm/submm radiation is absorbed. The absorption region is created by using a 1 mm long and 60 nm thick aluminum ($T_c = 1.2$ K) film for the center-strip. A thin-film niobium microstrip line then can couple the radiation into the aluminum. This microstrip to CPW coupling is used in the MKID camera design [9], but for the purpose of dark tests this microstrip line is not included on our device. An extra 54 $\mu$m length of Nb CPW inductor is inserted between the Nb IDC and the Al CPW inductor to control the resonance frequency. We have used an electromagnetic simulation software (Sonnet [10]) to extract equivalent parallel LC circuit parameters, the coupling quality factor ($Q_c$), and resonance frequency ($f_r$) for these resonators. The simulations show that the IDC is nearly a lumped element with $C = 1.92$ pF at the resonance frequency $f_r = 5.56$ GHz. A different type of lumped element interdigitated capacitor has been used in independent work [11].

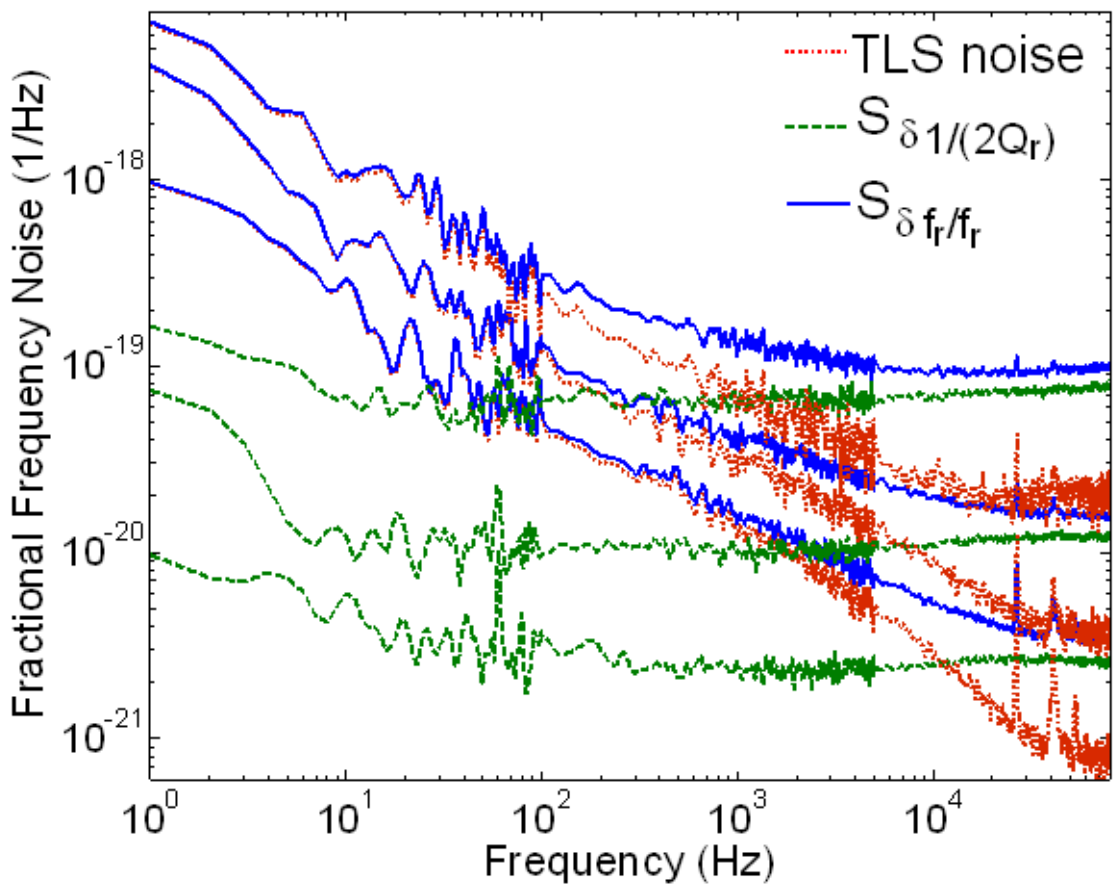

**FIGURE 3.** (Color online) Measured fractional frequency noise spectra (single-sided) for three different microwave readout powers. From top to bottom the three curves in each group correspond to $P_{\mu w}$ = -109, -103, and -97 dBm. For each power, the different noise components and the total noise are plotted as indicated in the legend.

## MEASUREMENT RESULTS

The device is cooled in a dilution refrigerator to a temperature of 120 mK. A microwave synthesizer is used to excite the resonator at $f_r = 5.56$ GHz. The transmitted signal through the feedline is amplified by a high electron mobility transistor (HEMT) amplifier (at 4 Kelvin) and is further amplified by a room temperature amplifier. An IQ mixer then compares this to the original signal from the synthesizer to produce I (in-phase) and Q (quadrature-phase) voltage outputs [1]. In order to keep the mixer properties unchanged as we change the readout power, a controllable attenuator is inserted in between the room temperature amplifier and mixer to keep the mixer RF port power constant. As we sweep the probe frequency around the resonance, a circle in the I-Q plane is formed. Time-domain noise data is taken by fixing the frequency to $f_r$ and recording the I and Q fluctuations. These are digitized with a sampling rate of 200 KHz. Anti-aliasing low-pass filters at 100 KHz are used. The center of the resonance loop is then subtracted from the data and then rotated around the origin such that the I and Q axes correspond to tangent and orthogonal directions to the circle at the resonance point. These directions correspond to fluctuations in the resonance frequency ($\delta f_r$) and dissipation ($\delta\, 1/Q_r$) of the resonator. Frequency-domain power spectra are calculated for the noise in each direction. The fractional frequency noise power spectra $S_{\delta f_r/f_r}(\nu)$ are plotted in Fig. 3 for three different microwave readout powers ($P_{\mu w}$). The dissipation direction spectra are nearly flat and correspond to the HEMT amplifier noise floor measured off resonance. They start to pick up 1/ν gain noise at $\nu < 10$ Hz as we increase the readout power. Since the amplifier noise contribution ($S_{\delta 1/2Q_r}(\nu)$), to the total frequency noise, is uncorrelated with the total frequency noise, we can subtract them to get the excess (TLS) frequency noise. The resulting spectrum ($S_{\delta f_r/f_r} - S_{\delta 1/2Q_r}$) indicated by dotted red lines in Fig. 3 has a typical TLS noise spectral shape and a ${P_{\mu w}}^{-1/2}$ power dependence, as universally observed in CPW resonators [4]. A key prediction of the noise model in [6] is that the noise originates in the capacitive portion of the resonator, rather than the inductive portion, and therefore the model can be tested by changing only the capacitive portion, leaving the inductive portion and the current passing through it the same. In our comparison, to take into account the difference in CPW center-strip widths, we use current density ($K$). The root mean square (rms) current density is calculated in units of A/m for the CPW and IDC cases as follows:

$$K^{CPW} = \frac{1}{w_{ind}} \left[8 P_{\mu w} {Q_r}^2 / \pi Q_c Z_r\right]^{1/2} \qquad (1)$$

$$K^{IDC} = \frac{1}{w_{ind}}\left[2\omega_r C P_{\mu w}\, Q_r{}^2/Q_c\right]^{1/2} \quad (2)$$

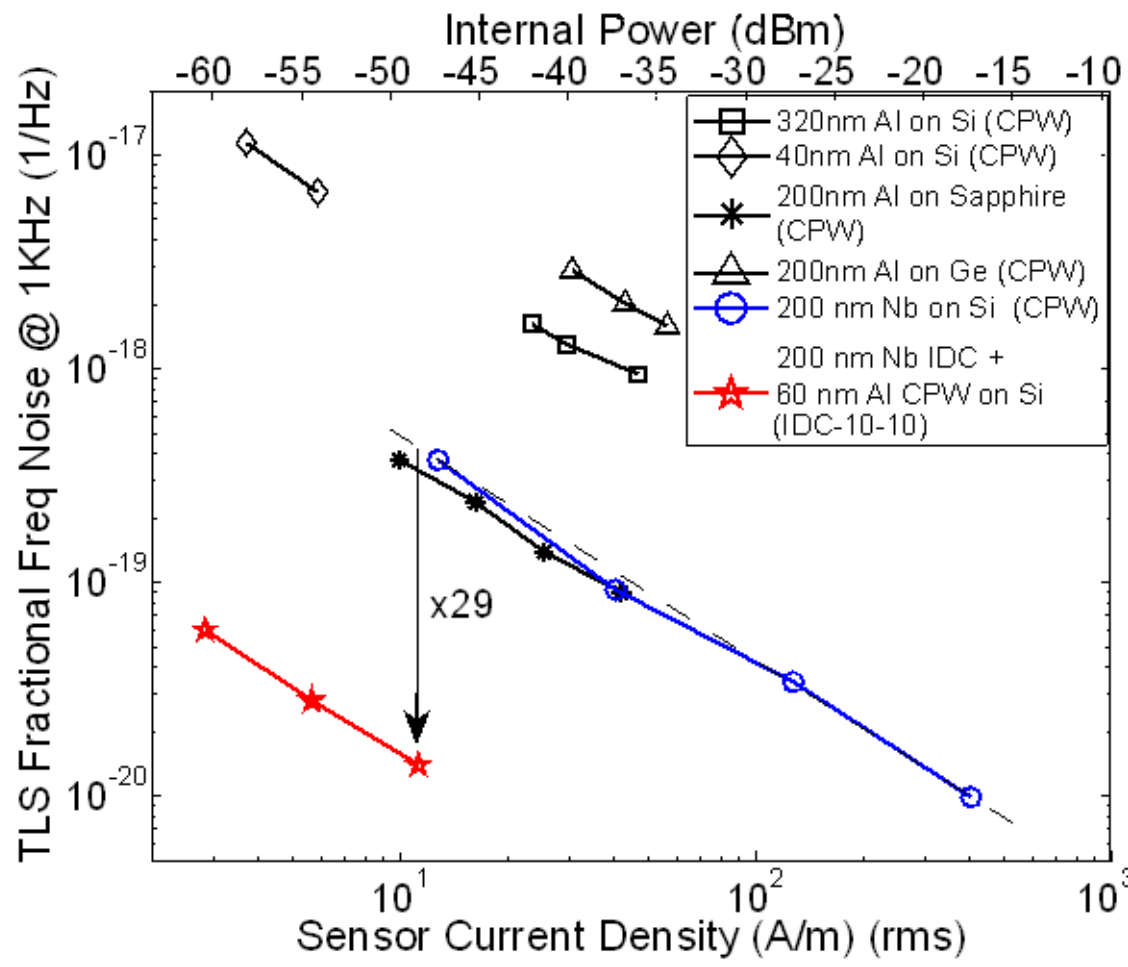

**FIGURE 4.** (Color online) Comparison of TLS fractional frequency noise measured at ν=1 KHz for CPW and IDC resonators. The bottom horizontal axis is the current density (K) flowing in the short circuit part of the resonator. The top horizontal axis is microwave power inside the resonator ($P_{int}$) and applies to CPW resonators only. The data for the first 5 resonators are from [4] and all have similar geometry CPW's ($w$= 3 $\mu$m, and $s$ = 2 $\mu$m). The data indicated by pentagon stars is for our IDC resonator with a $w$= 6 $\mu$m, and $s$ = 2 $\mu$m CPW inductor.

Here, $w_{ind}$ is the width of the CPW center strip in the inductive (sensory) end, $Z_r$ is the CPW characteristic impedance, $C$ is the IDC capacitance, $\omega_r$ is the resonance frequency, $Q_r$ is the resonance quality factor, $Q_c$ is the coupling quality factor, and $P_{\mu w}$ is the microwave readout power. We have rescaled the internal power ($P_{int} = 2Q_r{}^2 P_{\mu w}/\pi Q_c$) axis used for the CPW resonator data in [4] to sensor current density, and plotted the results along with the new IDC resonators in Fig. 4. The IDC resonators have ~29 times lower TLS noise compared to Nb CPW resonators. This confirms the hypothesis that the noise is predominantly generated in the capacitive portion, and is clearly against the hypothesis that the noise is coming from the inductive part of the CPW near the shorting end suggested by [7]. A direct numerical comparison to the noise scaling theory [6] would require exact knowledge of the $\vec{E}$ field distribution and exact TLS distribution in our IDC resonators which we do not have.

The noise equivalent power (NEP) for an optically unloaded IDC resonator can be calculated [1,2,3] using the measured noise spectra along with measured responsivities and lifetimes and is plotted in Fig. 5. We have obtained quasi-particle lifetimes of ~50 $\mu$s for our 60 nm aluminum films. With an IDC resonator design and longer lifetimes up to 2.3 ms reported by [12], dark frequency readout NEPs of ~1×$10^{-18}$ or below should be achievable.

## SUMMARY

Using the predictions provided by a semi empirical noise model [6] we have successfully designed and fabricated resonators with a new capacitive geometry that provide dramatic reduction by a factor of 29 in noise caused by two levels system (TLS) defects . This confirms that the noise is predominantly generated in the capacitive portion in our original CPW resonators. These new devices are replacing the CPW resonators in our next design iteration in progress for MKIDCam [9]. There are considerable prospects for further reducing or eliminating the TLS noise. For example, TLS physics predicts [13] that the noise should go down with resonator frequency. Therefore, using larger IDCs will result in lower noise (and improved multiplexing). A more aggressive method is to eliminate the TLS noise by using parallel-plate capacitors with a purely crystalline dielectric that is nearly TLS free. We are currently pursuing both these methods.

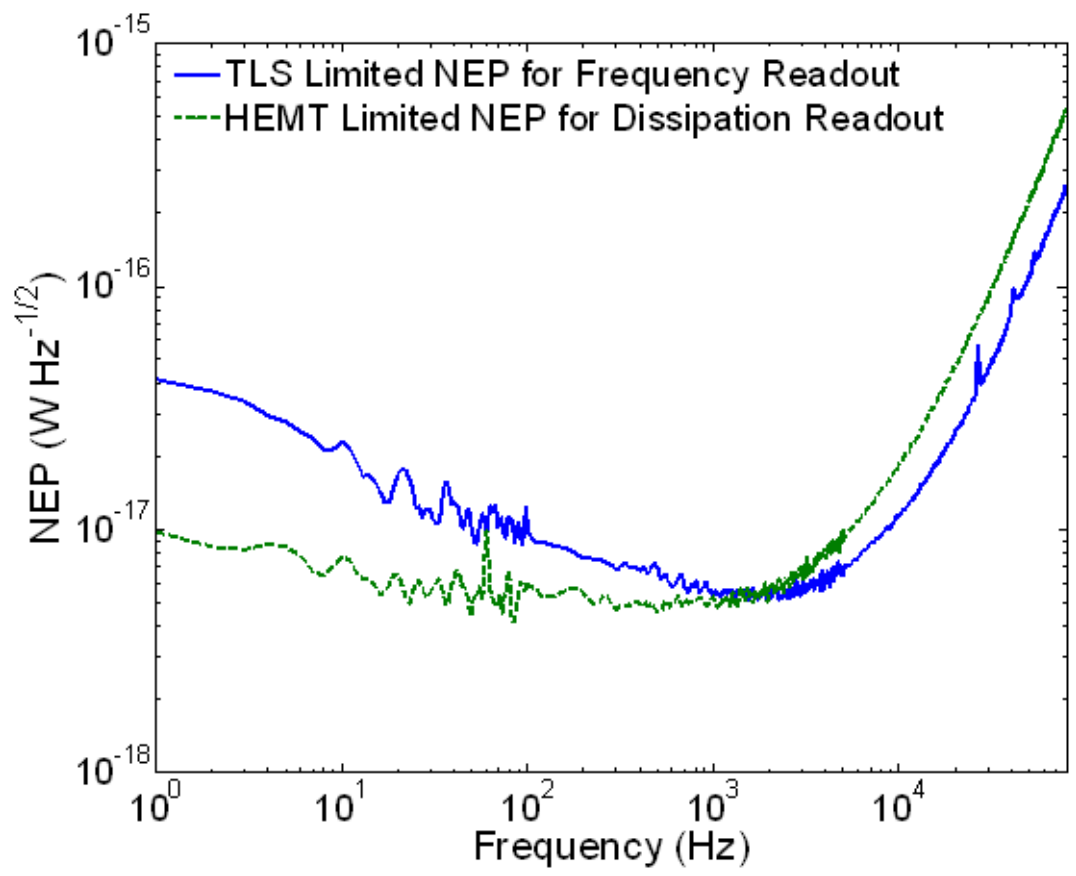

**FIGURE 5.** Dark NEP at 120 mK and $P_{\mu w}$ = -97 dBm with $\tau_{qp}$ = 50 $\mu$s for 60 nm Al films. Measured responsivities are $\delta f_r/f_r/\delta N_{qp} = 2.51\times10^{-11}, \delta\,(1/2Q_r)/\delta N_{qp} = 1.06\times10^{-11}$.